\documentclass[5p,times,number,sort&compress]{elsarticle}
\usepackage[utf8x]{inputenc}
\usepackage{fixltx2e}
\usepackage{microtype}
\usepackage{fix-cm}
\usepackage{ellipsis}
\usepackage{amsmath,amssymb}
\usepackage[colorlinks=false,pdfborder={0 0 0}]{hyperref}
\usepackage[amsmath,hyperref,thmmarks]{ntheorem}
\usepackage{asymptote}
\usepackage{booktabs}
\usepackage[table]{xcolor}
\usepackage[english]{babel}

\theoremstyle{plain}
\newtheorem{theorem}{Theorem}[section]
\newtheorem{lemma}[theorem]{Lemma}

\newtheorem{corollary}[theorem]{Corollary}
\newtheorem{definition}[theorem]{Definition}

\theoremstyle{nonumberplain}
\theoremheaderfont{\normalfont\itshape}
\theorembodyfont{\normalfont}
\theoremseparator{.}
\theoremsymbol{$\square$}
\newtheorem{proof}{Proof}

\def\allFormulae{\mathcal{L}}
\def\ie{\emph{i.e.}}
\def\eg{\emph{e.g.}}

\def\true{1}
\def\xor{\oplus}
\def\false{0}
\def\id{\mathrm{id}}
\newcommand\redTo[1][\calC]{\mathrel{\leq_{m}^{#1}}}
\newcommand\equivTo[1][\calC]{\mathrel{\equiv_{m}^{#1}}}
\def\imp{\mathrel{\rightarrow}}
\def\nimp{\mathrel{\nrightarrow}}
\newcommand\dual[1]{\mathrm{dual}(#1)}
\newcommand\NC[1]{\mathrm{NC}^{#1}}
\newcommand\AC[1]{\mathrm{AC}^{#1}}
\newcommand\BFMin{\mathrm{BFMIN}\xspace}
\newcommand\N{\mathbb{N}\xspace}
\newcommand\SigmaP[1]{{\ensuremath\Sigma^\mathrm{p}_{#1}}\xspace}

\def\calC{{\cal C}}

\def\CloneBF{\mathsf{BF}}
\def\CloneR{\mathsf{R}}
\def\CloneM{\mathsf{M}}
\def\CloneS{\mathsf{S}}
\def\CloneV{\mathsf{V}}
\def\CloneD{\mathsf{D}}
\def\CloneE{\mathsf{E}}
\def\CloneL{\mathsf{L}}
\def\CloneN{\mathsf{N}}
\def\CloneI{\mathsf{I}}

\makeatletter
\def\ps@pprintTitle{%
  \let\@oddhead\@empty
  \let\@evenhead\@empty
  \def\@oddfoot{\reset@font\hfil\thepage\hfil}
  \let\@evenfoot\@oddfoot
}
\makeatother

\begin{document}

\begin{frontmatter}

\title{On the Applicability of Post's Lattice\tnoteref{DFG}}
\author{Michael Thomas}
\address{%
  TWT~GmbH\\
  Bernh\"{a}user Stra\ss{}e 40--42\\
  73765 Neuhausen auf den Fildern
}
\tnotetext[DFG]{Work supported by DFG grant VO 630/6-2 and performed while employed at the Gottfried Wilhelm Leibniz Universit\"{a}t Hannover.\\
\indent\hphantom{\footnotemark[1]}
\textit{Email address:}\space\raggedright\texttt{michael.thomas@twt-gmbh.de} (or \texttt{thomas@thi.uni-hanover.de})}

\begin{abstract}
  \noindent
  For decision problems $\Pi(B)$ defined over Boolean circuits using gates from a restricted set $B$ only, 
  we have $\Pi(B) \redTo[\AC{0}] \Pi(B')$ for all finite sets $B$ and $B'$ of gates such that all gates from $B$ can be 
  computed by circuits over gates from $B'$.
  In this note, we show that a weaker version of this statement holds for decision problems defined over Boolean
  formulae, namely that $\Pi(B) \redTo[\NC2] \Pi(B' \cup \{\land,\lor\})$ and $\Pi(B) \redTo[\NC2] \Pi(B' \cup \{\false,\true\})$ 
  for all finite sets $B$ and $B'$ of Boolean functions such that all $f \in B$ can be defined in $B'$.
\end{abstract}


\end{frontmatter}
\section{Introduction}

Let $\Pi$ denote some decision problem defined over Boolean circuits 
such that membership in $\Pi$ is invariant under 
the substitution of equivalent circuits. 
Denote by $\Pi(B)$ its restriction to circuits using gates from a finite set $B$ only. 
It is easily observed that then $\Pi(B) \redTo[\AC{0}] \Pi(B')$ for all finite sets $B$ and $B'$ 
such that all gates from $B$ can be computed by circuits over gates from $B'$ (see, \eg,
\cite{boehler-creignou-reith-vollmer:03,reith-wagner:00}).
If we consider formulae instead, this reduction does not necessarily hold; 
the size of the smallest formula over the Boolean connectives from $B'$ 
computing some function from $B$ might be of exponential size.

Building on works of \cite{spira:71,br74,bobu94},
we show that a weaker form of this property holds for decision problems defined over formulae, namely that 
$\Pi(B) \redTo[\NC2] \Pi(B' \cup \{\land,\lor\})$ and $\Pi(B) \redTo[\NC2] \Pi(B' \cup \{\false,\true\})$ 
for all finite sets $B$ and $B'$ of Boolean functions such that all $f \in B$ can be defined in $B'$. 
Moreover, 
if all connectives in $B$ can be expressed using either only conjunction ($\land$), only disjunction ($\lor$) or 
only the exclusive-or ($\xor$), we obtain $\Pi(B) \redTo[\AC{0}] \Pi(B')$, as in the circuit setting.

These results provide a (partial) account for the polytomous complexity classifications 
of problems parametrized by the set of available Boolean connectives:
the complexity of the satisfiability problem was, for instance, shown to be
$\mathrm{NP}$-complete if $x \nimp y$  can be composed from the available Boolean connectives, 
and solvable in logspace in all other cases~\cite{lewis:79}. 
Further results include a variety of problems in 
  propositional logic~\cite{reith:03,beyersdorff-meier-thomas-vollmer:09a},
  modal logics~\cite{bauland-hemaspaandra-schnoor-schnoor:06}, 
  temporal logics~\cite{bauland-schneider-schnoor-schnoor-vollmer:08,bauland-mundhenk-schneider-schnoor-schnoor-vollmer:09,meier-mundhenk-thomas-vollmer:08,beyersdorff-meier-mundhenk-schneider-thomas-vollmer:09}, 
  their hybrid variants~\cite{meier-mundhenk-schneider-thomas-weber-weiss:09,meier-mundhenk-schneider-thomas-weiss:10}, and 
  nonmonotonic logics~\cite{thomas:09,beyersdorff-meier-thomas-vollmer:09b,creignou-meier-thomas-vollmer:10}.
  
We point out that the results obtained herein are completely general in that they do not rely on properties of the considered problems except invariance of membership under substitution of logically equivalent formulae (\ie, if $(\varphi,x)$ is an instance of $\Pi$ with $\varphi$ being a Boolean formula and if $\varphi'$ is a Boolean formula logically equivalent to $\varphi$, then $(\varphi,x) \in \Pi$ iff $(\varphi',x) \in \Pi$). This generality comes at the price of a fairly powerful reduction. However, in practice, most problems exhibit additional structure that allow to further restrict the notion of reductions considered.

\section{Preliminaries}

\paragraph{Propositional Logic}
Let $\allFormulae$ be the set of propositional formulae, \ie, 
the set of formulae defined via
\[
  \varphi ::= a \mid c(\varphi,\ldots,\varphi), 
\]
where $a$ is a proposition and $c$ is an $n$-ary connective.
We associate an $n$-ary connective $c$ with the $n$-ary Boolean function $f_c\colon\{\false,\true\}^n \to \{\false,\true\}$ defined by
$f(a_1,\ldots,a_n):=\true$ if and only if the formula $c(x_1,\ldots,x_n)$ becomes true when assigning 
$a_i$ to $c_i$, $1 \leq i \leq n$.
Let $\varphi_{[\alpha / \beta]}$ denote $\varphi$ with all occurrences of the subformula $\alpha$ replaced by some formula $\beta$.
For a finite set $B$ of Boolean connectives, 
let $\allFormulae(B)$ denote the set of \emph{$B$-formulae}, \emph{i.e.}, the set $\allFormulae$ restricted to formulae using connectives from $B$ only.
The depth of a formula is the maximum nesting depth of Boolean connectives; 
the size of a formula is equal to the number of symbols used to represent it.

\paragraph{Clones and Post's Lattice}
A \emph{clone}\index{clone} is a set of Boolean functions that is closed under superposition, \emph{i.e.},
$B$ contains all projections (the functions $f(x_1,\ldots,x_n) = x_k$ for all $1 \leq k \leq n$) 
and is closed under arbitrary composition \cite{pippenger:97}. 
For a set $B$ of Boolean functions, we denote by $[B]$\index{$[\cdot]$} the smallest clone containing $B$ and 
call $B$ a \emph{base}\index{base} for $[B]$. 
A $B$-formula $g$ is called \emph{$B$-representation}\index{B-representation@$B$-representation} of $f$ if $f$ and $g$ are equivalent, \emph{i.e.}, $f \equiv g$.
It is clear that $B$-representations exist for every $f \in [B]$. 

In \cite{post:41}, Post showed that the set of all clones 
ordered by inclusion together with $[A \cap B]$ and $[A \cup B]$ forms a lattice
and found a finite base for each clone, see Figure~\ref{fig:post's_lattice}.
To introduce the clones, we define the following properties. 
Say that a set $A \subseteq \{\false,\true\}^n$ is \emph{$c$-separating}, $c \in \{\false,\true\}$, 
if there exists an $i \in \{1, \ldots , n\}$ such that $(a_1,\ldots,a_n) \in A$ implies $a_i = c$. 
Let $f$ be an $n$-ary Boolean function and define the dual of $f$ to be the Boolean function 
$\dual{f}(x_1, \ldots , x_n) := \neg f(\neg x_1, \ldots , \neg x_n)$.
We say that 
\begin{itemize} \itemsep 0pt
  \item $f$ is \emph{$c$-reproducing} if $f(c, \ldots , c) = c$, $c \in \{\false,\true\}$;
  \item $f$ is \emph{$c$-separating} if $f^{-1}(c)$ is $c$-separating, $c \in \{\false,\true\}$;
  \item $f$ is \emph{$c$-separating of degree $m$} if all $A \subseteq f^{-1}(c)$ with $|A|=m$ are $c$-separating;
  \item $f$ is \emph{monotone} if $a_1 \leq b_1, a_2 \leq b_2, \ldots , a_n \leq b_n$ implies $f(a_1, \ldots , a_n) \leq f(b_1, \ldots , b_n)$;
  \item $f$ is \emph{self-dual} if $f \equiv \dual{f}$;
  \item $f$ is \emph{affine} if $f(x_1,\ldots,x_n) \equiv x_1 \xor \cdots \xor x_n \xor c$ with $c \in \{\false,\true\}$;
  \item $f$ is \emph{essentially unary} if $f$ depends on at most one variable.
\end{itemize}
The above properties canonically extend to sets $B$ of Boolean functions by requiring that all $f \in B$ satisfy the given property. 
The list of all clones is given in Table~\ref{tab:clones}. 

\begin{table*}
  \small
  \setlength{\aboverulesep}{0pt}
  \setlength{\belowrulesep}{1pt}
  \rowcolors{2}{gray!10}{}
  \centering
  \begin{tabular}{lll}
  \toprule
  Clone & Definition & Base \\
  \midrule
  $\CloneBF$    & All Boolean functions & $\{x \land y ,\neg x\}$ \\
  $\CloneR_0$   & $\{f \in \CloneBF \mid f \text{ is $\false$-reproducing}\}$ & $\{ x \land y , x \xor y \}$ \\
  $\CloneR_1$   & $\{f \in \CloneBF \mid f \text{ is $\true$-reproducing}\}$ & $\{x \lor y, x \leftrightarrow y \}$ \\
  $\CloneR_2$   & $\CloneR_0 \cap \CloneR_1$ & $\{x \lor y, x \land (y \leftrightarrow z) \}$ \\
  $\CloneM$     & $\{f \in \CloneBF \mid f \text{ is monotone}\}$ & $\{x \land y, x\lor y,\false,\true\}$ \\
  $\CloneM_0$   & $\CloneM \cap \CloneR_0$ & $\{x \land y,x\lor y,\false\}$ \\
  $\CloneM_1$   & $\CloneM \cap \CloneR_1$ & $\{x \land y,x\lor y,\true\}$ \\  
  $\CloneM_2$   & $\CloneM \cap \CloneR_2$ & $\{x \land y,x\lor y\}$ \\
  $\CloneS_0$   & $\{f \in \CloneBF \mid f \text{ is $\false$-separating}\}$ & $\{x \imp y\}$ \\
  $\CloneS_0^n$ & $\{f \in \CloneBF \mid f \text{ is $\false$-separating of degree $n$}\}$ & $\{x \imp y,\dual{\mathrm{t}^{n+1}_n}\}$ \\
  $\CloneS_1$   & $\{f \in \CloneBF \mid f \text{ is $\true$-separating}\}$ & $\{x \nimp y\}$ \\
  $\CloneS_1^n$ & $\{f \in \CloneBF \mid f \text{ is $\true$-separating of degree $n$}\}$ & $\{x \nimp y,\mathrm{t}^{n+1}_n\}$ \\
  $\CloneS_{02}^n$& $\CloneS_0^n \cap \CloneR_2$ & $\{x \lor (y \land \neg z), \dual{\mathrm{t}^{n+1}_n}\}$ \\
  $\CloneS_{02}$  & $\CloneS_0 \cap \CloneR_2$ & $\{x \lor (y \land \neg z)\}$ \\
  $\CloneS_{01}^n$& $\CloneS_0^n \cap \CloneM$ & $\{\dual{\mathrm{t}^{n+1}_n},\true\}$ \\
  $\CloneS_{01}$  & $\CloneS_0 \cap \CloneM$ & $\{x \lor (y \land z),\true\}$ \\
  $\CloneS_{00}^n$& $\CloneS_0^n \cap \CloneR_2 \cap \CloneM$ & $\{x \lor (y \land z),\dual{\mathrm{t}^{n+1}_n}\} $ \\
  $\CloneS_{00}$  & $\CloneS_0 \cap \CloneR_2 \cap \CloneM$ & $\{x \lor (y \land z)\}$ \\
  $\CloneS_{12}^n$& $\CloneS_1^n \cap \CloneR_2$ & $\{x\land (y\lor \neg z),\mathrm{t}^{n+1}_n\}$ \\
  $\CloneS_{12}$  & $\CloneS_1 \cap \CloneR_2$ & $\{x\land (y\lor \neg z)\}$ \\
  $\CloneS_{11}^n$& $\CloneS_1^n \cap \CloneM$ & $\{\mathrm{t}^{n+1}_n,\false\}$ \\
  $\CloneS_{11}$  & $\CloneS_1 \cap \CloneM$ & $\{x\land (y\lor z),\false\}$ \\
  $\CloneS_{10}^n$& $\CloneS_1^n \cap \CloneR_2 \cap \CloneM$ & $\{x\land (y\lor z),\mathrm{t}^{n+1}_n\}$ \\
  $\CloneS_{10}$  & $\CloneS_1 \cap \CloneR_2 \cap \CloneM$ & $\{x\land (y\lor z)\}$ \\
  
  $\CloneD$     & $\{f \in \CloneBF \mid f \text{ is self-dual}\}$ & $\{(x \!\land\! y) \lor (x\! \land\! \neg z) \lor (\neg y \!\land \!\neg z)\}$ \\
  $\CloneD_1$   & $\CloneD \cap \CloneR_2$ & $\{(x \!\land \!y) \lor (x \!\land \!\neg z) \lor (y \!\land \!\neg z)\}$ \\
  $\CloneD_2$   & $\CloneD \cap \CloneM$ & $\{(x\! \land \!y) \lor (x\! \land \!z) \lor (y \!\land \!z)\}$ \\  
  
  $\CloneL$     & $\{f \in \CloneBF \mid f \text{ is affine}\}$ & $\{x \xor y,\true\}$ \\
  $\CloneL_0$   & $\CloneL \cap \CloneR_0$ & $\{x \xor y\}$ \\
  $\CloneL_1$   & $\CloneL \cap \CloneR_1$ & $\{x \leftrightarrow y\}$ \\
  $\CloneL_2$   & $\CloneL \cap \CloneR_2$ & $\{x \xor y \xor z\}$ \\
  $\CloneL_3$   & $\CloneL \cap \CloneD$ & $\{x \xor y \xor z \xor \true \}$ \\
  
  $\CloneE$     & $\{f \in \CloneBF \mid f \text{ is constant or a conjunction}\}$ & $\{x \land y,\false, \true\}$ \\
  $\CloneE_0$   & $\CloneE \cap \CloneR_0$ & $\{x \land y,\false\}$ \\
  $\CloneE_1$   & $\CloneE \cap \CloneR_1$ & $\{x \land y,\true\}$ \\
  $\CloneE_2$   & $\CloneE \cap \CloneR_2$ & $\{x \land y\}$ \\
  
  $\CloneV$     & $\{f \in \CloneBF \mid f \text{ is constant or a disjunction}\}$ & $\{x \lor y,\false,\true\}$ \\
  $\CloneV_0$   & $\CloneV \cap \CloneR_0$ & $\{x \lor y,\false\}$ \\
  $\CloneV_1$   & $\CloneV \cap \CloneR_1$ & $\{x \lor y,\true\}$ \\
  $\CloneV_2$   & $\CloneV \cap \CloneR_2$ & $\{x \lor y\}$ \\
  
  $\CloneN$     & $\{f \in \CloneBF \mid f \text{ is essentially unary}\}$ & $\{\neg x,\false,\true\}$ \\ 
  $\CloneN_2$   & $\CloneN \cap \CloneD$ & $\{\neg x\}$ \\

  $\CloneI$     & $\{f \in \CloneBF \mid f \text{ is constant or a projection}\}$ & $\{\id,\false,\true\}$ \\ 
  $\CloneI_0$   & $\CloneI \cap \CloneR_0$ & $\{\id,\false\}$ \\
  $\CloneI_1$   & $\CloneI \cap \CloneR_1$ & $\{\id,\true\}$ \\
  $\CloneI_2$   & $\CloneI \cap \CloneR_2$ & $\{\id\}$ \\
  \bottomrule
  \end{tabular}
  \caption{List of all clones with definition and bases,
  where $\id$ denotes the identity and 
  $\mathrm{t}^{n+1}_n(x_0,\ldots,x_n) := \bigvee_{i=0}^{n} (x_0\land \cdots \land x_{i-1} \land x_{i+1} \land \cdots \land x_n)$.}
  \label{tab:clones}
\end{table*}

\begin{figure*}
\centering
\begin{asy}
  import lattice;
  Lattice lattice = Lattice(1cm, 0.75cm, 0.3cm);
  lattice.draw();
\end{asy}
\caption{\label{fig:post's_lattice}Post's lattice}
\end{figure*}

\paragraph{Reductions}

Let $A$ and $B$ be decision problems.
Say that $A$ \emph{$\calC$ many-one reduces} to $B$ (written: $A \redTo B$) if there exists a $\calC$-computable function $f$
mapping instances $x$ of $A$ to instances $f(x)$ of $B$ such that $x \in A \iff f(x) \in B$.
If $A \redTo B$ and $B \redTo A$, we also write $A \equivTo B$.

\section{Previous Results and Auxiliary Lemmas}

The following lemma due to Spira is well-known and will be useful if the given set of Boolean functions is functionally complete.

\begin{lemma}[{\cite{spira:71}}] \label{lem:spira}
  Let $\varphi$ be a propositional formula.
  Then there exists an equivalent $\{\land,\lor,\neg\}$-formula $\psi$ such that
  the depth of $\psi$ is $O(\log |\varphi|)$ and the size of $\psi$ is $|\varphi|^{O(1)}$.
\end{lemma}

%

\begin{lemma} \label{lem:logdepth-S00-formula}
  Let $\varphi$ be a propositional formula over Boolean connectives from $[B] \subseteq \CloneM$ and 
  let $g(x,y,z):= x \lor (y \land z)$.
  Then there exists an equivalent $(B \cup \{g,\false,\true\})$-formula $\psi$ such that
  the depth of $\psi$ is $O(\log |\varphi|)$ and the size of $\psi$ is $|\varphi|^{O(1)}$.
\end{lemma}
\begin{proof}
  We proceed analogous to a construction of Bonet and Buss from~\cite{bobu94}.
  Let $\varphi$ be the given formula over connectives from a set $B$ and let 
  $m$ be the number of occurrences of propositions in $\varphi$.
  We claim that there exists an equivalent $(B \cup \{g,\false,\true\})$-formula 
  of depth $O(\log m)$ and polynomial size.
  
  If $m \leq 1$ then $\varphi$ is equivalent to $x$ or a constant and can be implemented in depth $1$.
  Hence assume that $m > 1$ and that the claim holds for all smaller $m$.
  Then there exists a subformula $\psi$ that contains $\geq \frac{m}k$ occurrences of propositions,
  where $k$ is a bound on the arity of the functions in $B$ (see also \cite{br74}).
  Define $\varphi':=g(\varphi_{[\psi/\false]},\varphi_{[\psi/\true]},\psi) 
  \equiv \varphi_{[\psi/\false]} \lor (\varphi_{[\psi/\true]} \land \psi).$
  By monotonicity, $\varphi$ is equivalent to $\varphi'$.
  Moreover, by induction hypothesis, we may assume the depths of $\psi$ and $\varphi_{[\psi/c]}$, $c \in \{\false,\true\}$, 
  to be $O\big(\log \frac{m}k\big)$ and $O\big(\log \frac{(k-1)m}{k}\big)$, respectively.
	Denote by $d$ the constant hidden in these $O$-notations.
  Then the depth of $\varphi'$ can be bounded by 
  $2+\max\{\mathrm{depth}(\varphi_{[\psi/\false]}),\linebreak[1] \mathrm{depth}(\varphi_{[\psi/\true]}),\linebreak[1] \mathrm{depth}(\psi)\} 
  = 2 + d \cdot k \cdot\log \big(\frac{(k-1)m}k\big) 
  = 2 + d \cdot k \cdot\Big(\log  m + \log\big(1 - \frac1k\big)\Big) 
  < 2 + d \cdot k \cdot\big(\log  m -\frac1k \big) 
  \in O(\log m)$, as $\log\big(1-\frac1k\big)<-\frac1k$.
  Concluding, the size of $\varphi'$ is at most quadratic in the size of $\varphi$.
\end{proof}

\begin{lemma} \label{lem:logdepth-S10-formula}
  Let $\varphi$ be a propositional formula over Boolean connectives from $B \subseteq \CloneM$ and 
  let $h(x,y,z):= x \land (y \lor z)$.
  Then there exists an equivalent $(B \cup \{h,\false,\true\})$-formula $\psi$ such that
  the depth of $\psi$ is $O(\log |\varphi|)$ and the size of $\psi$ is $|\varphi|^{O(1)}$.
\end{lemma}
\begin{proof}
  Analogous to Lemma~\ref{lem:logdepth-S00-formula} using 
  $\varphi' \kern-2.5pt := h(\varphi_{[\psi/\true]},\varphi_{[\psi/\false]},\psi) 
  \equiv \varphi_{[\psi/\true]} \land (\varphi_{[\psi/\false]} \lor \psi)$ in the inductive step.
\end{proof}

\section{Results}

Throughout this section, 
let $B$ and $B'$ be for finite sets of Boolean connectives and $\Sigma$ be an alphabet.
We will first formalize the notion of problems defined over propositional formulae and
invariance under the substitution of equivalent $B$-formulae. 

\begin{definition}
  A \emph{decision problem defined over (propositional) formulae} is any set of $\Pi \subseteq \Sigma^\star \times \allFormulae$. 
  We will write $\Pi(B)$ for $\Pi \cap \big(\Sigma^\star \times \allFormulae(B)\big)$.
  
  Further, say that a decision problem $\Pi(B)$ defined over propositional formulae is 
  \emph{invariant under the substitution of equivalent formulae} if 
  $(\varphi,x) \in \Pi$ if and only if $(\psi,x) \in \Pi$ for all formulae $\psi$ equivalent to $\varphi$.
\end{definition}

\begin{lemma} \label{lem:belowEVL}
	Fix $B$ and let $\Pi(B)$ be a decision problem defined over propositional formulae that is invariant under the substitution of equivalent formulae. Then the following holds for all $B'$ satisfying $B \subseteq [B']$:
  \begin{enumerate}
		\item If $[B] \subseteq \CloneE$ or $[B] \subseteq \CloneV$, then $\Pi(B) \redTo[\AC0] \Pi(B')$.
		\item If $[B] \subseteq \CloneL$, then $\Pi(B) \redTo[{\AC0[2]}] \Pi(B')$.
  \end{enumerate}
\end{lemma}
\begin{proof}
	First suppose that $[B] \subseteq \CloneE$ and let $\Pi(B)$ be as in the statement of the lemma
  Then any $B$-formula $\varphi$ over propositions $x_1,x_2,\ldots$ is equivalent to a formula $\varphi' := c \land \bigwedge_{i \in I} x_i$, where $c \in \{\false,\true\}$. This representation is computable in logarithmic space, as 
  $c = \false$ iff $\varphi$ is not satisfied by the assignment setting all propositions to $\true$ (\ie, $\varphi(\true,\ldots,\true)=\false$), and
  $i \in I$ iff $\varphi(\true,\ldots,\true)=\true$ and $\varphi$ is not satisfied by the assignment setting all propositions but $x_i$ to $\true$.
  By inserting parentheses, $\varphi'$ can be transformed into a formula of logarithmic depth such that replacing all occurring constants and connectives with their $B'$-representations yields an equivalent $B'$-formula $\varphi''$ of size at most $2^{O(\log |\varphi|)} = |\varphi|^{O(1)}$.
  Thus, given input $(\varphi,x) \in \allFormulae(B) \times \Sigma^\star$, it suffices to construct $(\varphi',x)$.
  As the evaluation of $B$-formulae for $[B] \subseteq \CloneE$ can be performed in $\AC0$~\cite{schnoor:10}, we finally obtain 
$\Pi(B) \redTo[\AC0] \Pi(B')$ for all $B'$ satisfying $B \subseteq [B']$.

  For $[B] \subseteq \CloneL$ and $[B] \subseteq \CloneV$, similar arguments work. 
  The construction of $\varphi' := c \xor \bigoplus_{i \in I} x_i$ (resp.\ $\varphi' := c \lor \bigvee_{i \in I} x_i$) is as follows:
  $c \equiv \true$ iff $\varphi(\false,\ldots,\false)=\true$, and
  $i \in I$ iff the truth value of $\varphi$ under the assignment setting all propositions to $\false$ and the truth value of $\varphi$ under the assignment setting only the proposition $x_i$ to $\true$ differ
  (resp.\ $i \in I$ iff $\varphi(\false,\ldots,\false)=\false$ and $\varphi$ is satisfied by the assignment setting only the proposition $x_i$ to $\true$).
  And the evaluation of $B$-formulae for $[B] \subseteq \CloneV$ can be performed in $\AC{0}$, while for $[B] \subseteq \CloneL$ we require $\AC{0}[2]$.
\end{proof}

Henceforth, let $\calC \supseteq \AC0$ be such that given $\varphi$ the formula $\psi$ in the Lemmas~\ref{lem:spira},~\ref{lem:logdepth-S00-formula} and~\ref{lem:logdepth-S10-formula} can be computed in $\calC$. (A direct implementation of these restructurings requires $O(\log^2 n)$ space, 
hence $\NC2 \subseteq \calC$ suffices; Cook and Gupta showed that Spira's construction can actually be performed in 
alternating $O(\log n \cdot \log \log n)$-time~\cite{gupta:85}).

\begin{lemma} \label{lem:aboveS00/S10}
	Fix $B$ and let $\Pi(B)$ be a decision problem defined over propositional formulae that is invariant under the substitution of equivalent formulae. Then the following holds for all $B'$ satisfying $B \subseteq [B']$:
	\begin{enumerate}
		\item If $\CloneS_{00} \subseteq [B] \subseteq \CloneM$,
		then $\Pi(B) \redTo \Pi(B' \cup \{\land\})$.
		\item 
		If $\CloneS_{10} \subseteq [B] \subseteq \CloneM$,
		then $\Pi(B) \redTo \Pi(B' \cup \{\lor\})$.
	\end{enumerate}
\end{lemma}
\begin{proof}
	Suppose that $\CloneS_{00} \subseteq [B] \subseteq \CloneM$ and let $\Pi(B)$ be as in the statement of the lemma.
	Let $(\varphi,x)$ be the given instance with $\varphi \in \allFormulae(B)$.
  Denote by $g(x,y,z)$ the function $x \lor (y \land z) \in \CloneS_{00} \subseteq [B]$.
  Then, by Lemma~\ref{lem:logdepth-S00-formula},
  there exists a $(B \cup \{g,\false,\true\})$-formula $\varphi'$ of logarithmic depth and polynomial size such that $\varphi \equiv \varphi'$.
  Obtain $\varphi'$ from $\varphi$ by replacing all connectives from $B \cup \{g\}$ with their $B'$-representations.
  Next, if $\true \notin [B']$, we eliminate the constant $\true$ by replacing it with the $B'$-representation of $\bigvee_{i=1}^n x_i$,
  where $x_1,\ldots,x_n$ enumerate all propositions occurring in $\varphi'$.
  Analogously, if $\false \notin [B']$, we eliminate the constant $\false$ by replacing it with the $B'$-representation of $\bigwedge_{i=1}^n x_i$.  
  Call the resulting formula $\varphi''$.
  If $\true \notin [B']$, then $\varphi$ cannot be satisfied by the assignment setting all propositions to $\false$,
  as $[B'] \subseteq \CloneR_0$;
  for all other assignments, $\bigvee_{i=1}^n x_i$ is satisfied. 
  If $\false \notin [B']$, then $\varphi$ is satisfied by the assignment setting all propositions to $\true$,
  as $[B] \subseteq \CloneR_1$;
  for all other assignments, $\bigwedge_{i=1}^n x_i$ is not satisfied.
  Therefore, $\varphi''$ is equivalent to $\varphi$.
  
  Consequently, the mapping $(\varphi,x) \mapsto (\varphi'',x)$ constitutes a $\redTo$-reduction from $\Pi(B)$ to $\Pi(B' \cup \{\land\})$, as $\varphi'$ is $\calC$-computable by assumption and the construction of $\varphi''$ from $\varphi'$ requires local replacements only. This concludes the proof of the first claim. 
  
  As for the second claim, suppose that $\CloneS_{10} \subseteq [B] \subseteq \CloneM$.
  Denote again by $(\varphi,x)$ the given instance and abbreviate with 
	$h(x,y,z)$ the function $x \land (y \lor z) \in \CloneS_{10} \subseteq [B]$.
	By Lemma~\ref{lem:logdepth-S10-formula},
  there exists a $(B \cup \{h,\false,\true\})$-formula $\varphi'$ of logarithmic depth and polynomial size such that $\varphi \equiv \varphi'$.
  Obtain $\varphi''$ from $\varphi$ by replacing all connectives from $B \cup \{h\}$ 
  with their $B'$-representations and eliminating the constants not contained in $[B']$ as above.
  Then $(\varphi,x) \mapsto (\varphi'',x)$ constitutes a $\redTo$-reduction from $\Pi(B)$ to $\Pi(B' \cup \{\lor\})$. 
\end{proof}

\begin{lemma} \label{lem:aboveS02/S12}
  Fix $B$ and let $\Pi(B)$ be a decision problem defined over propositional formulae that is invariant under the substitution of equivalent formulae. Then the following holds for all $B'$ satisfying $B \subseteq [B']$:
	\begin{enumerate}
		\item If $\CloneS_{02} \subseteq [B]$,
		then $\Pi(B) \redTo \Pi(B' \cup \{\land\})$.
		\item 
		If $\CloneS_{12} \subseteq [B]$,
		then $\Pi(B) \redTo \Pi(B' \cup \{\lor\})$.
	\end{enumerate}
\end{lemma}
\begin{proof}
	Suppose that $\CloneS_{02} \subseteq [B]$ and let $\Pi(B)$ be as in the statement of the lemma.
	Let $(\varphi,x)$ be the given instance with $\varphi \in \allFormulae(B)$.
  By Lemma~\ref{lem:spira}, there exists a $\{\land,\lor,\neg\}$-formula $\varphi'$ of logarithmic depth 
  and polynomial size such that $\varphi \equiv \varphi'$.
  Observe that $\varphi'$ can be constructed from $\varphi$ by a procedure 
  similar to that used in the proof of Lemma~\ref{lem:logdepth-S00-formula} (in the inductive step, use
  $(\varphi_{[\psi/\false]} \land \neg \psi) \lor (\varphi_{[\psi/\true]} \land \psi)$ as the new formula).
  As $x \lor (y \land \neg z)$ is a base for $[B']$ and 
  $x \lor (y \land \neg \false) \equiv x \lor y$,
  $\false \lor (y \land \neg (\false \lor (\true \land \neg z))) \equiv y \land z$ and
  $\false \lor (\true \land \neg z) \equiv \neg z$,
  we obtain $\{\land,\lor,\neg\} \in [B' \cup \{\false,\true\}]$.
  So we can first replace all connectives from $B \cup \{\land,\lor,\neg\}$ in 
  $\varphi'$ with their $(B' \cup \{\false,\true\})$-representations, and 
  second, eliminate those constants not contained in $[B']$ as in the proof of 
  Lemma~\ref{lem:aboveS00/S10}.
  Call the resulting formula $\varphi''$.
  As $\CloneS_{02} \subseteq [B']$ and $\true \notin [B']$ imply that $[B'] \subseteq \CloneR_0$,
  and $\CloneS_{02} \subseteq [B']$ and $\false \notin [B']$ imply that $[B'] \subseteq \CloneR_1$,
  $\varphi''$ is equivalent to $\varphi'$ by the same arguments as above.
  The function mapping $(\varphi,x)$ to $(\varphi'',x)$ is hence a $\redTo$-reduction from $\Pi(B)$ to $\Pi(B' \cup \{\land\})$. 
  
  The proof of the second claim is analogous.
\end{proof}

\begin{lemma} \label{lem:D2-to-D}
	Fix $B$ and let $\Pi(B)$ be a decision problem defined over propositional formulae that is invariant under the substitution of equivalent formulae. 
	If $\CloneD_{2} \subseteq [B] \subseteq \CloneD$, 
	then $\Pi(B) \redTo \Pi(B' \cup \{\lor\})$ and $\Pi(B) \redTo \Pi(B' \cup \{\land\})$ for all $B'$ satisfying $B \subseteq [B']$.
\end{lemma}
\begin{proof}
	Let $B$ and $\Pi(B)$ be as in the statement of the lemma and denote by $(\varphi,x)$ the given instance with $\varphi \in \allFormulae(B)$.
  On the one hand, if $[B]=\CloneD_2$, then by Lemma~\ref{lem:logdepth-S00-formula} and Lemma~\ref{lem:logdepth-S10-formula} there exist logarithmic-depth polynomial-size formulae $\varphi' \in \allFormulae(B \cup \{x \lor (y \land z),\false,\true\})$ and $\varphi'' \in \allFormulae(B \cup \{x \land (y \lor z),\false,\true\})$. Proceeding as in the proof of Lemma~\ref{lem:aboveS00/S10}, we obtain the desired reduction.
  
  On the other hand, if $\CloneD_1 \subseteq [B]$, then by Lemma~\ref{lem:spira} there exists a $\{\land,\lor,\neg\}$-formula $\varphi'$ of logarithmic depth 
  and polynomial size such that $\varphi \equiv \varphi'$.
  As $[B' \cup \{\false,\true\}] = \CloneBF$,
  we may replace all connectives in $\varphi'$ with their $(B' \cup \{\false,\true\})$-representations.
  If $[B'] \subseteq \CloneR_0$ (or if $[B'] \subseteq \CloneR_1$), we may eliminate the constant $\true$ (or $\false$) as in the proof of Lemma~\ref{lem:aboveS00/S10}. Otherwise, if $[B']=\CloneBF$, then we may replace $\true$ with $t \lor \neg t$ and $\false$ with $t \land \neg t$, where $t$ is an arbitrary fresh proposition.
  Either way, we obtain a formula $\varphi'' \in \allFormulae(B' \cup C)$ of polynomial size such that $\varphi'' \equiv \varphi$ and $C$ is either $\{\lor\}$, $\{\land\}$, or the empty set.
  The mapping from $(\varphi,x) \in \Pi(B)$ to $(\varphi',x) \in \Pi(B' \cup C)$ is the desired $\redTo$-reduction.
\end{proof}

We are now ready to state our main theorem.

\begin{theorem} \label{thm:prob(b)<=prob(b'+and/or)}
	Fix $B$ and let $\Pi(B)$ be a decision problem defined over propositional formulae that is invariant under the substitution of equivalent formulae. Then the following holds for all $B'$ satisfying $B \subseteq [B']$:
  \begin{itemize}\itemsep 0pt
    \item If $[B] \subseteq \CloneV$ or $[B] \subseteq \CloneL$ or $[B] \subseteq \CloneE$ or $\CloneM_2 \subseteq [B]$, 
    then $\Pi(B) \redTo \Pi(B')$.
    \item If $\CloneS_{00} \subseteq [B] \subseteq \CloneS_0^2$ or $\CloneD_2 \subseteq [B] \subseteq \CloneD$,
    then $\Pi(B) \redTo \Pi(B' \cup \{\land\})$.
    \item If $\CloneS_{10} \subseteq [B] \subseteq \CloneS_1^2$ or $\CloneD_2 \subseteq [B] \subseteq \CloneD$,
    then $\Pi(B) \redTo \Pi(B' \cup \{\lor\})$.
  \end{itemize}
\end{theorem}
\begin{proof}
	Consider the lattice in Fig.~\ref{fig:post's_lattice}.
	It holds that either 
    (a) $[B] \subseteq \CloneV$, 
    (b) $[B] \subseteq \CloneL$, 
    (c) $[B] \subseteq \CloneE$,
    (d) $\CloneS_{00} \subseteq [B] \subseteq \CloneS_{0}^{2}$, 
    (e) $\CloneS_{10} \subseteq [B] \subseteq \CloneS_{1}^{2}$,
    (f) $\CloneD_2 \subseteq [B] \subseteq \CloneD$, or
    (g) $\CloneM_2 \subseteq [B]$.
  The first claim corresponds to the cases (a)--(c) and (g).
  The second and third claim correspond to case (d) and (f) resp.\ (e) and (f).
  
  In cases (a)--(c), $\Pi(B) \redTo \Pi(B')$ follows from Lemma~\ref{lem:belowEVL}.

  As for case (d), 
  we have either $[B] \subseteq \CloneS_{01}^{2}$ or $\CloneS_{02} \subseteq [B]$. 
  In either case, the reduction $\Pi(B) \redTo \Pi(B' \cup \{\land\})$ is implied by 
  Lemmas~\ref{lem:aboveS00/S10} and~\ref{lem:aboveS02/S12}.

	Case (e) analogously yields $\Pi(B) \redTo \Pi(B' \cup \{\lor\})$.

  For case (f), Lemma~\ref{lem:D2-to-D} yields both $\Pi(B) \redTo \Pi(B' \cup \{\land\})$ and $\Pi(B) \redTo \Pi(B' \cup \{\lor\})$.

  It remains to consider case (g):
  Fix a set $B$ with $\CloneM_2 \subseteq [B]$.
  If we suppose that $[B] \subseteq \CloneM$, Lemma~\ref{lem:aboveS00/S10} yields $\Pi(B) \redTo \Pi(B' \cup \{\land\})$ (or $\Pi(B) \redTo \Pi(B' \cup \{\lor\})$). Yet, for all such $B$, we have $\{\land,\lor\} \in [B] \subseteq [B']$; in which case Lemma~\ref{lem:aboveS00/S10} actually yields $\Pi(B) \redTo \Pi(B')$. The same argument applies if $\CloneM \nsubseteq [B]$, using Lemma~\ref{lem:aboveS02/S12} instead.
  This completes the last case and establishes the theorem.
\end{proof}

As an easy consequence of Theorem~\ref{thm:prob(b)<=prob(b'+and/or)} and the remark below Lemma~\ref{lem:belowEVL},
we obtain the following two corollaries:

\begin{corollary}\label{cor:prob(b)<=prob(b')}
  If $\Pi(B \cup \{\false,\true\}) \redTo \Pi(B)$ for all $B$, 
  then $\Pi(B) \redTo \Pi(B')$ for all $B$ and $B'$ such that $B \subseteq [B']$;
  in particular, $\Pi(B)$ is $\calC$-equivalent to $\Pi$ restricted to one of the following sets of functions:
  $\{\land,\lor,\neg\},\{\land,\lor\},\{\land\},\{\lor\},\{\xor\},\{\neg\},\{\id\}$.
\end{corollary}

%

\begin{corollary}\label{cor:examples-bases}
  Let $B$ be a finite set of Boolean functions.
  \begin{itemize}\itemsep 0pt
    \item If $[B] = \CloneBF$, then $\Pi(B) \equivTo[\NC2] \Pi(\{\land,\lor,\neg\})$.
    \item If $[B] = \CloneM$, then $\Pi(B) \equivTo[\NC2] \Pi(\{\land,\lor,\false,\true\})$.
    \item If $[B] = \CloneL$, then $\Pi(B) \equivTo[{\AC{0}[2]}] \Pi(\{\xor,\true\})$.
    \item If $[B] = \CloneN$, then $\Pi(B) \equivTo[{\AC{0}[2]}] \Pi(\{\neg,\true\})$.
    \item If $[B] = \CloneE$, then $\Pi(B) \equivTo[\AC{0}] \Pi(\{\land,\false,\true\})$.
    \item If $[B] = \CloneV$, then $\Pi(B) \equivTo[\AC{0}] \Pi(\{\lor,\false,\true\})$.
  \end{itemize}
\end{corollary}

It is straightforward to extend Corollary~\ref{cor:examples-bases} to those clones not containing both constants.



\section{Concluding Remarks}

The results presented in this note provide insight into 
why complexity classifications of problems 
in Post's lattice yield only a finite number of complexity degrees. 

These results are completely general in the sense that
we did not place any restrictions on the considered decision problems $\Pi$ 
(unless, of course, that membership in $\Pi$ is invariant under substitution of equivalent formulae).
However, typically instances of natural decision problems exhibit additional structure; 
by exploiting this structure one may further reduce the computational power of the reduction $\redTo$, 
or obtain $\Pi(B) \redTo \Pi(B')$ without resorting to the assumption 
$\Pi(B \cup \{\false,\true\}) \redTo \Pi(B')$ given Corollary~\ref{cor:prob(b)<=prob(b')}.
For example, if $\Pi(\{\land,\lor\}) \redTo \Pi(B)$ for all finite sets $B$ of Boolean functions satisfying $\CloneS_{00} \subseteq [B]$ or $\CloneS_{10} \subseteq [B]$ or $\CloneD_{2} \subseteq [B]$, then $\Pi(B) \redTo \Pi(B')$ for all $B$ and $B'$ satisfying $B \subseteq [B']$. This holds for the propositional implication problem~\cite{beyersdorff-meier-thomas-vollmer:09a}, among others. 

It is worth noting that, on the other hand, there exist natural problems that do not satisfy the conditions imposed on $\Pi$ above.
Amongst those is the problem $\BFMin$, which asks to determine, given a Boolean formula and an integer $k$, whether there exists an equivalent formula of size $\leq k$. This problem has recently been shown to be $\SigmaP2$-complete for the Boolean standard base $B=\{\land,\lor,\neg\}$ using Turing reductions~\cite{buchfuhrer-umans:11}. However, considering its restriction to $B$-formulae, we obtain $\BFMin(B) \not\redTo \BFMin(B')$: 
Let $\varphi \notin \allFormulae(B)$ be some Boolean formula of arity $n$.
Then $(\varphi,c(n)) \in \BFMin(B \cup \{\varphi\})$ for some constant $c$ (depending on $n$ only), while 
$(\varphi,k) \notin \BFMin(B)$ for all $k \in \N$.

\bibliographystyle{plain}
\bibliography{posts-lattice}

\end{document}